\documentclass[letter]{aa}

\usepackage{graphicx}
\usepackage{txfonts}
\usepackage[colorlinks,citecolor=blue,urlcolor=blue,linkcolor=blue]{hyperref}
\usepackage{jab}

\providecommand{\RS}{\ensuremath{R_\odot}}
\providecommand{\rA}{\ensuremath{r_\mathrm{A}}}
\providecommand{\rE}{\ensuremath{r_\mathrm{E}}}
\providecommand{\sigmac}{\ensuremath{\sigma_\mathrm{c}}}
\providecommand{\sigmar}{\ensuremath{\sigma_\mathrm{r}}}
\providecommand{\phiHCS}{\ensuremath{\phi_\mathrm{HCS}}}

\begin{document} 
\title{The Near-Sun Streamer Belt Solar Wind: Turbulence and Solar Wind Acceleration}
\authorrunning{Chen et al.}
\author{C. H. K. Chen\inst{1}, B. D. G. Chandran\inst{2}, L. D. Woodham\inst{3}, S. I. Jones-Mecholsky\inst{4}, J. C. Perez\inst{5}, S. Bourouaine\inst{5,6}, T. A. Bowen\inst{7}, K. G. Klein\inst{8}, M. Moncuquet\inst{9}, J. C. Kasper\inst{10,11}, S. D. Bale\inst{1,3,7,12}}
\institute{School of Physics and Astronomy, Queen Mary University of London, London E1 4NS, UK\\\email{christopher.chen@qmul.ac.uk}
\and
Department of Physics and Astronomy, University of New Hampshire, Durham, NH 03824, USA
\and
Department of Physics, Imperial College London, London SW7 2AZ, UK
\and
NASA Goddard Space Flight Center, Greenbelt, MD 20771, USA
\and
Department of Aerospace, Physics and Space Sciences, Florida Institute of Technology, Melbourne, FL 32901, USA
\and
Johns Hopkins University Applied Physics Laboratory, Laurel, MD 20723, USA
\and
Space Sciences Laboratory, University of California, Berkeley, CA 94720, USA
\and
Lunar and Planetary Laboratory, University of Arizona, Tucson, AZ 85719, USA
\and
LESIA, Observatoire de Paris, Universit\'{e} PSL, CNRS, Sorbonne Universit\'{e}, Universit\'{e} de Paris, 92195 Meudon, France
\and
Climate and Space Sciences and Engineering, University of Michigan, Ann Arbor, MI 48109, USA
\and
Smithsonian Astrophysical Observatory, Cambridge, MA 02138 USA
\and
Physics Department, University of California, Berkeley, CA 94720, USA
}
\date{\today}
\abstract{
The fourth orbit of \emph{Parker Solar Probe} (\emph{PSP}) reached heliocentric distances down to 27.9\,\RS, allowing solar wind turbulence and acceleration mechanisms to be studied \emph{in situ} closer to the Sun than previously possible. The turbulence properties were found to be significantly different in the inbound and outbound portions of \emph{PSP}'s fourth solar encounter, likely due to the proximity to the heliospheric current sheet (HCS) in the outbound period. Near the HCS, in the streamer belt wind, the turbulence was found to have lower amplitudes, higher magnetic compressibility, a steeper magnetic field spectrum (with spectral index close to --5/3 rather than --3/2), a lower Alfv\'enicity, and a ``$1/f$'' break at much lower frequencies. These are also features of slow wind at 1 au, suggesting the near-Sun streamer belt wind to be the prototypical slow solar wind. The transition in properties occurs at a predicted angular distance of $\approx4^\circ$ from the HCS, suggesting $\approx8^\circ$ as the full-width of the streamer belt wind at these distances. While the majority of the Alfv\'enic turbulence energy fluxes measured by \emph{PSP} are consistent with those required for reflection-driven turbulence models of solar wind acceleration, the fluxes in the streamer belt are significantly lower than the model predictions, suggesting that additional mechanisms are necessary to explain the acceleration of the streamer belt solar wind.
}
\keywords{solar wind -- Sun: heliosphere -- plasmas -- turbulence -- waves}
\maketitle

\section{Introduction}

One of the major open questions in heliophysics is how the solar wind is accelerated to the high speeds measured \emph{in situ} by spacecraft in the solar system \citep{fox16}. Early models of solar wind generation, based on the pioneering work of \citet{parker58a}, were able to reproduce the qualitative properties of the solar wind, although could not explain all of the measured quantities seen at 1 au \citep[see, e.g., the reviews of][]{parker65,leer82,barnes92,hollweg08,hansteen12,cranmer15,cranmer17}. Turbulence is now thought to be one of the key processes playing a role in solar wind acceleration, providing both a source of energy to heat the corona \citep{coleman68} and a wave pressure to directly accelerate the wind \citep{alazraki71,belcher71b}. Possible mechanisms for the driving of the turbulence include reflection of the outward-propagating Alfv\'en waves by the large-scale gradients \citep{heinemann80,velli93} and velocity shears \citep{coleman68,roberts92}. Models that incorporate these effects are now able to reproduce most solar wind conditions at 1 au \citep[e.g.,][]{cranmer07,verdini10,chandran11,vanderholst14,usmanov18,shoda19}, but more stringent tests come from comparing their predictions to measurements close to the Sun.

The nature of the turbulence in the solar wind and plasma turbulence in general is also a major open question  \citep{bruno13,alexandrova13a,kiyani15,chen16b}. Initial results from \emph{Parker Solar Probe} (\emph{PSP}) have revealed many similarities, but also some key differences in the near-Sun solar wind turbulence. Both the power levels and cascade rates were found to be several orders or magnitude larger at $\sim36\,\RS$ compared to 1 au \citep{bale19,chen20,bandyopadhyay20}. The turbulence was also found to be less magnetically compressible, more imbalanced, with a shallower magnetic field spectral index of $\approx-3/2$. The low compressibility and polarisation is consistent with a reduced slow mode component to the turbulence \citep{chen20,chaston20}. Both the outer scale and ion scale spectral breaks move to larger scales approximately linearly with heliocentric distance \citep{chen20,duan20}, indicating that the width of the MHD inertial range stays approximately constant over this distance range. The steep ion-scale transition range, however, is more prominent closer to the Sun, indicating stronger dissipation or increase of the cascade rate \citep{bowen20c}, or perhaps a build-up of energy at these scales \citep{meyrand20}. The overall increase of the turbulence energy flux, compared to the bulk solar wind kinetic energy flux, was found by \citet{chen20} to be consistent with the reflection-driven turbulence solar wind model of \citet{chandran11}, showing that this remains a viable mechanism to explain the acceleration of the open field wind. Other comparisons of the early \emph{PSP} data to turbulence-driven models also report agreement \citep{bandyopadhyay20,reville20a,adhikari20b}. 

Much of the solar wind measured in the early \emph{PSP} solar encounters has been of open-field coronal hole origin \citep{bale19,panascenco20,badman20a,badman20b}, although short periods of streamer belt wind near the heliospheric current sheet (HCS) were also identified \citep{szabo20,rouillard20,lavraud20}. Encounter 4, however, was different in that for the majority of the outbound portion, \emph{PSP} was consistently in streamer belt plasma \citep{bale20}. In this Letter, the properties of turbulence during this encounter are presented. These are compared to the distance to the HCS to show the differences between the streamer belt wind and the open field wind. The turbulence energy flux is also compared to predictions of reflection-driven turbulence solar wind models to investigate the acceleration mechanisms of the streamer belt wind.

\section{Data}

\begin{figure}
\centering
\includegraphics[width=\columnwidth]{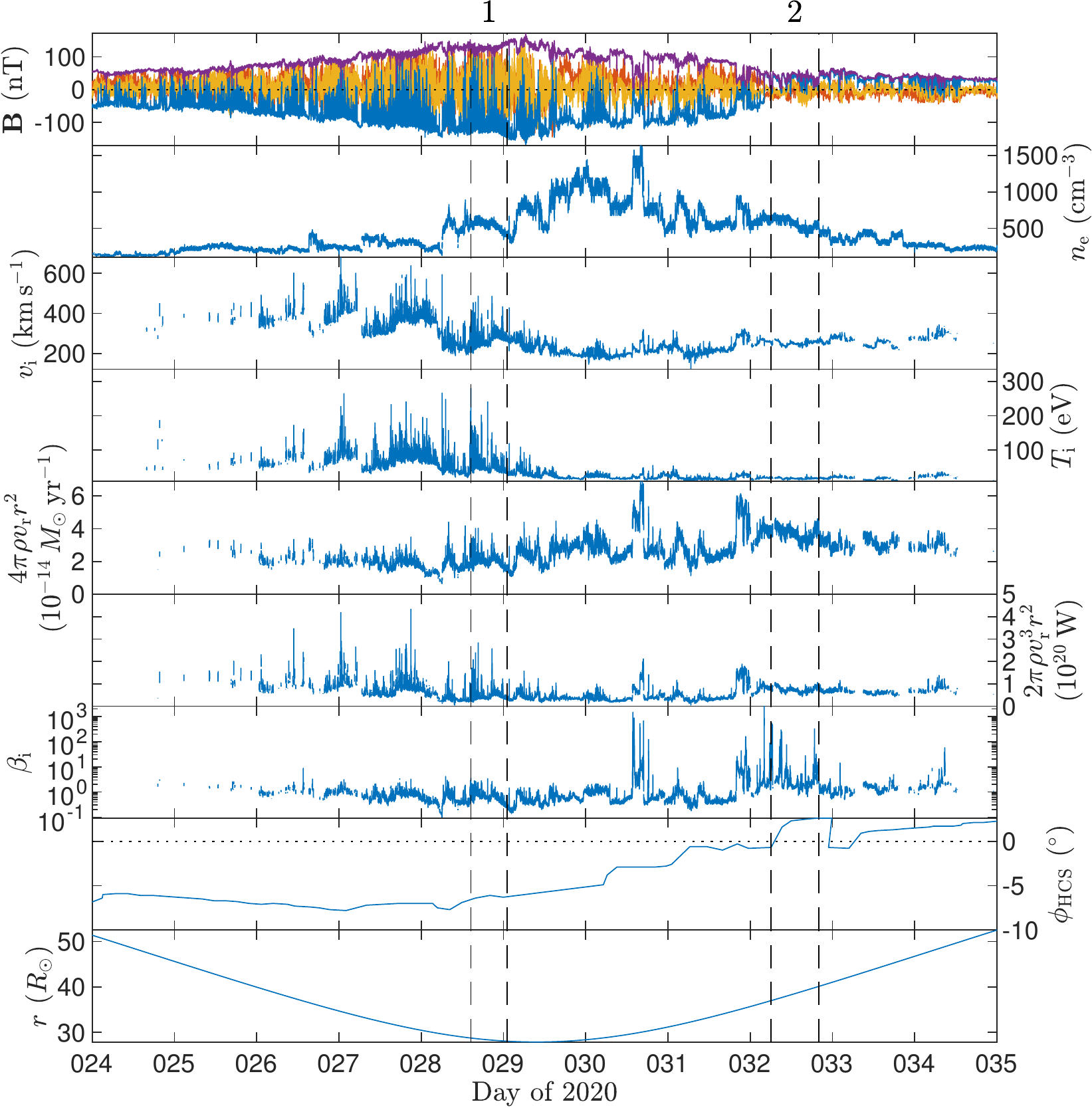}
\caption{Time series of Encounter 4 solar wind parameters, with symbols defined in the text. \phiHCS\ is the predicted angular distance of the spacecraft to the heliospheric current sheet. The vertical dashed lines indicate the intervals used to calculate the spectra in Figure \ref{fig:spectra}.}
\label{fig:timeseries}
\end{figure}

Data from \emph{PSP} \citep{fox16}, primarily from its 4th solar encounter, but also from all of the first four orbits, were used for this study. Magnetic field, $\mathbf{B}$, and electron density, $n_\mathrm{e}$, were obtained from the MAG and RFS/LFR instruments of the FIELDS suite \citep{bale16}; the electron density measurement is described in \citet{moncuquet20}. Ion (proton) velocity, $\mathbf{v}_\mathrm{i}$, and temperature, $T_\mathrm{i}$ were obtained primarily from from the SPAN-I \citep{livi20}, but also the SPC \citep{case20}, instruments of the SWEAP suite \citep{kasper16}. The SPAN-I data consist of bi-Maxwellian fits to the proton core population, described in \citet{woodham20}, with the same selection  criteria used for excluding bad fits from the dataset, and this data is used for $\mathbf{v}_\mathrm{i}$ and $T_\mathrm{i}$ unless stated otherwise. Since the fluctuations investigated in this Letter are at MHD scales, the solar wind velocity $\mathbf{v}$ is taken to be equal to $\mathbf{v}_\mathrm{i}$.

\begin{figure}
\centering
\includegraphics[width=\columnwidth,trim=0 0 0 0,clip]{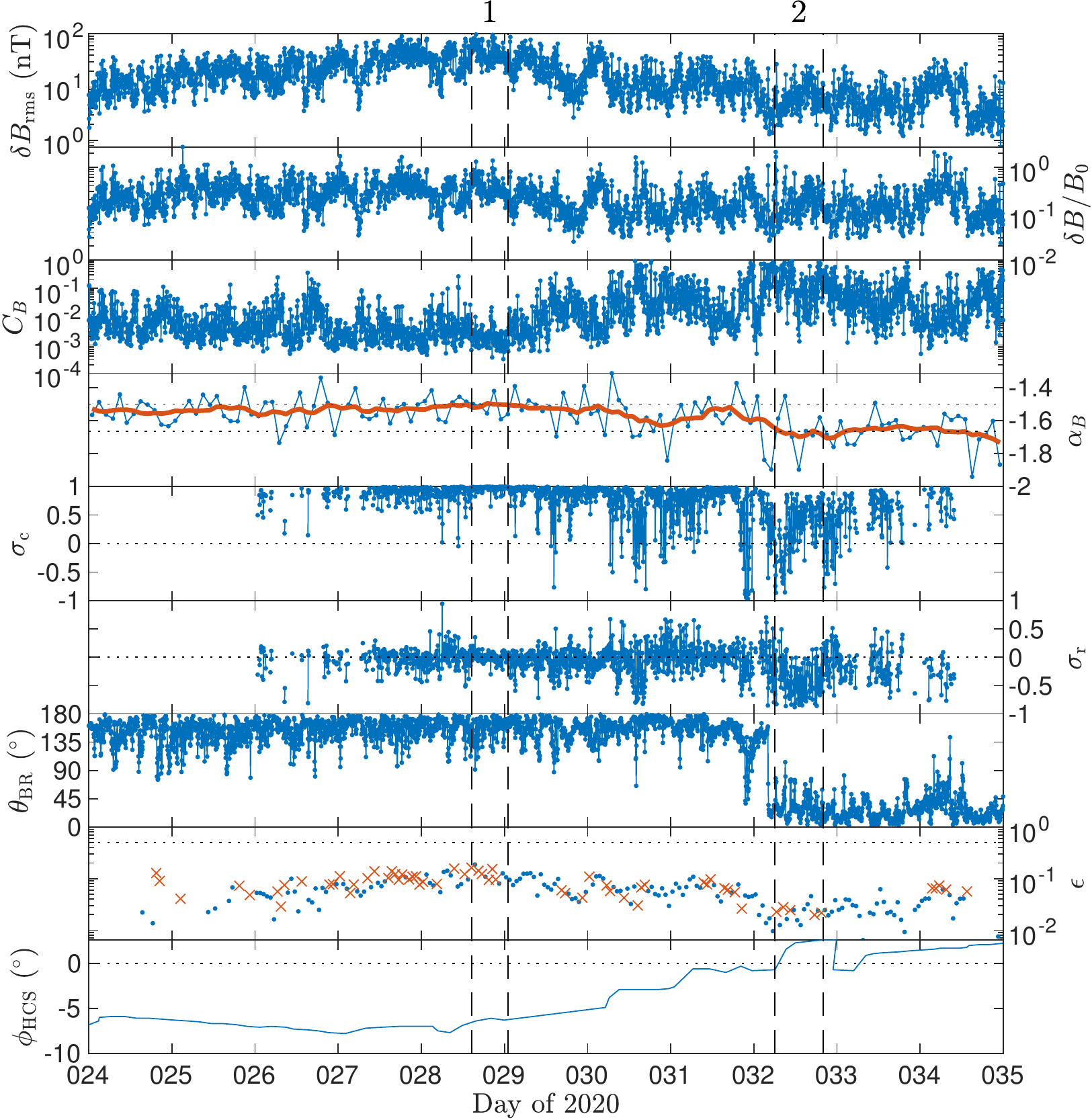}
\caption{Time series of Encounter 4 turbulence properties, with symbols defined in the text. The red line is a 10-point running mean of $\alpha_B$. The vertical dashed lines indicate the intervals used to calculate the spectra in Figure \ref{fig:spectra}.}
\label{fig:turbulencetimeseries}
\end{figure}

A time series of the data for Encounter 4 is shown in Figure \ref{fig:timeseries}. Additional quantities plotted include the distance-normalised mass flux, $4\pi\rho v_\mathrm{r}r^2$, where $\rho$ is the total mass density estimated as $\rho=m_\mathrm{p}n_\mathrm{e}(1+3f_\alpha)/(1+f_\alpha)$, where $f_\alpha=0.05$ is the assumed alpha fraction of the ion number density, $v_\mathrm{r}$ is the radial solar wind speed and $r$ is the radial distance of the spacecraft to the Sun, the distance-normalised kinetic energy flux, $2\pi\rho v_\mathrm{r}^3r^2$, and the ion plasma beta, $\beta_\mathrm{i}=2\mu_0n_\mathrm{i}k_\mathrm{B}T_\mathrm{i}/B^2$. \phiHCS\ is the angle (centred at the Sun) of the spacecraft to the heliospheric current sheet (HCS) estimated using the Wang-Sheeley-Arge (WSA) model, which consists of a PFSS model for the inner corona encased in a Schatten current sheet shell \citep{arge00,arge03,arge04,szabo20}.  A zero-point corrected GONG synoptic magnetogram was chosen for the model's inner boundary condition, which was found to produce solar wind predictions that correspond well with the IMF polarity inversion observed by \emph{PSP} near the end of January. It can be seen that, unlike the previous encounters, the inbound and outbound portions had different solar wind properties: the outbound period had higher density, lower speed and temperature, with a higher mass flux and plasma beta. These differences can be accounted for by the fact that \emph{PSP} spent most of the outbound period close to the HCS.

\section{Results}
\subsection{Turbulence Properties}

\begin{figure}
\centering
\includegraphics[width=\columnwidth,trim=0 0 0 0,clip]{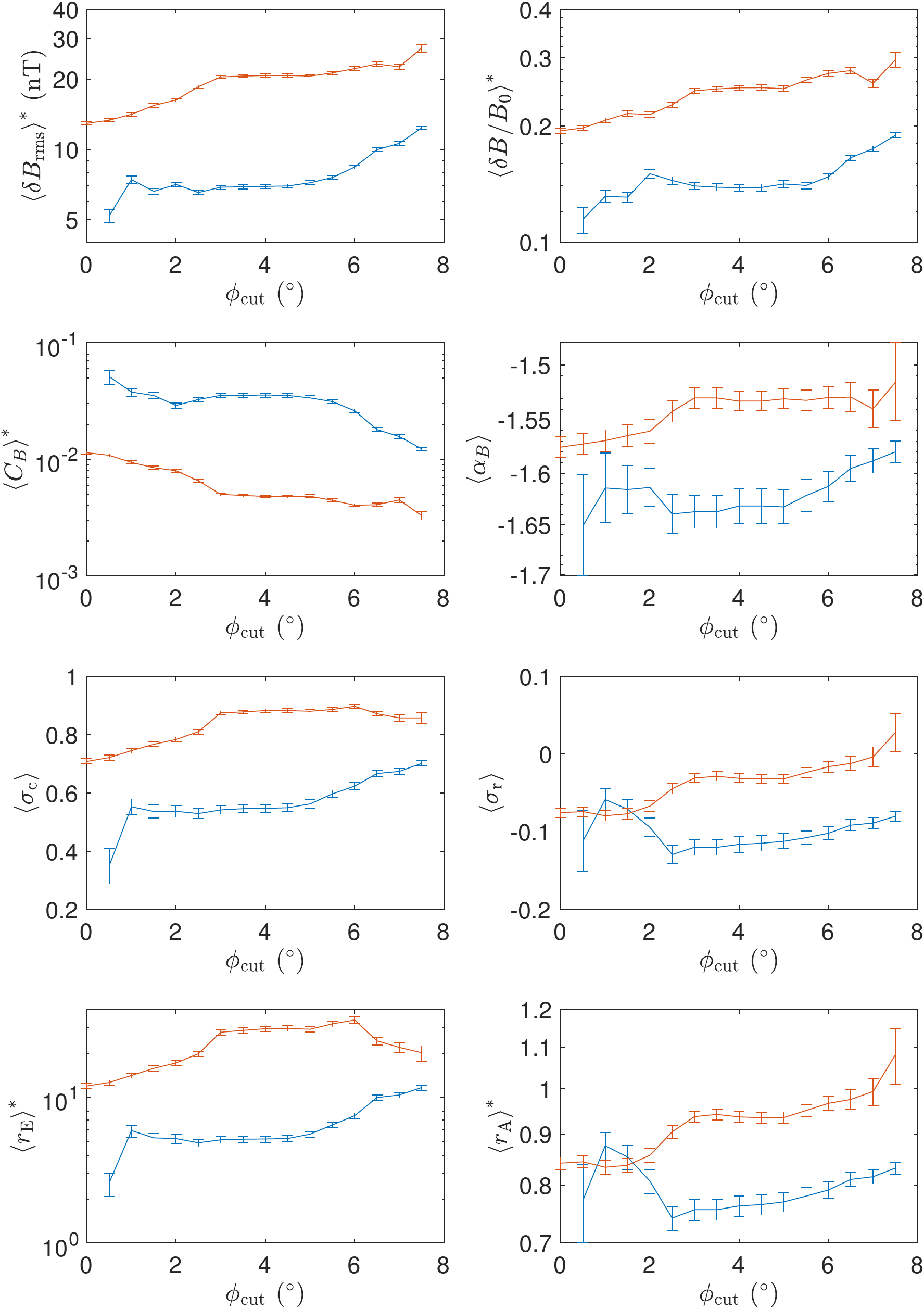}
\caption{Average turbulence properties for times close to (blue) and far from (red) the HCS as a function of the value $\phi_\mathrm{cut}$ used to define close and far. Averages are arithmetic means except for quantities marked with $^*$, which are geometric means. The error bars represent the standard error of the mean. The largest overall difference in average values is at $\phi_\mathrm{cut}\approx4^\circ$.}
\label{fig:cutvalues}
\end{figure}

The 11-day period of Encounter 4 (days 24-34 of 2020) was divided into intervals of 300\,s duration, roughly comparable to the outer scale \citep{chen20,parashar20,bandyopadhyay20,bourouaine20b}, and in each a set of turbulence properties was calculated: the total rms magnetic fluctuation amplitude
\begin{equation}
\delta B_\mathrm{rms}=\sqrt{\left<|\delta\mathbf{B}|^2\right>},
\end{equation}
where $\delta\mathbf{B}=\mathbf{B}-\mathbf{B}_0$, $\mathbf{B}_0=\left<\mathbf{B}\right>$, and the angular brackets denote a time average, in this case over each 300\,s interval, the normalised rms fluctuation amplitude $\delta B/B_0$, the magnetic compressibility, 
\begin{equation}
C_B=\sqrt{\frac{\left<(\delta|\mathbf{B}|)^2\right>}{\left<|\delta\mathbf{B}|^2\right>}}
\end{equation}
the normalised cross helicity, 
\begin{equation}
\sigmac=\frac{2\left<\delta\mathbf{v}\cdot\delta\mathbf{b}\right>}{\left<|\delta\mathbf{v}|^2+|\delta\mathbf{b}|^2\right>},
\end{equation}
where $\mathbf{b}=\mathbf{B}/\sqrt{\mu_0\rho_0}$ and $\delta\mathbf{v}=\mathbf{v}-\left<\mathbf{v}\right>$, the normalised residual energy, 
\begin{equation}
\sigmar=\frac{2\left<\delta\mathbf{z}^+\cdot\delta\mathbf{z}^-\right>}{\left<|\delta\mathbf{z}^+|^2+|\delta\mathbf{z}^-|^2\right>},
\end{equation}
where the Elsasser fields are $\delta\mathbf{z}^\pm=\delta\mathbf{v}\pm\delta\mathbf{b}$, and the angle between the magnetic field and the radial direction, 
\begin{equation}
\theta_\mathrm{BR}=\cos^{-1}\left(\hat{\mathbf{B}}_0\cdot\hat{\mathbf{r}}\right).
\end{equation}
In the definition of $\mathbf{b}$, its sign is reversed if $\theta_\mathrm{BR}<90^\circ$ so that positive \sigmac\ corresponds to Alfv\'enic propagation away from the Sun. In addition, the MHD inertial range magnetic field spectral index, $\alpha_B$, was calculated from the FFT of 2-hour intervals and fitting a power-law function in the range of spacecraft-frame frequencies $10^{-2}\,\mathrm{Hz}<f_\mathrm{sc}<10^{-1}\,\mathrm{Hz}$.

\begin{figure}
\centering
\includegraphics[width=\columnwidth,trim=0 0 0 0,clip]{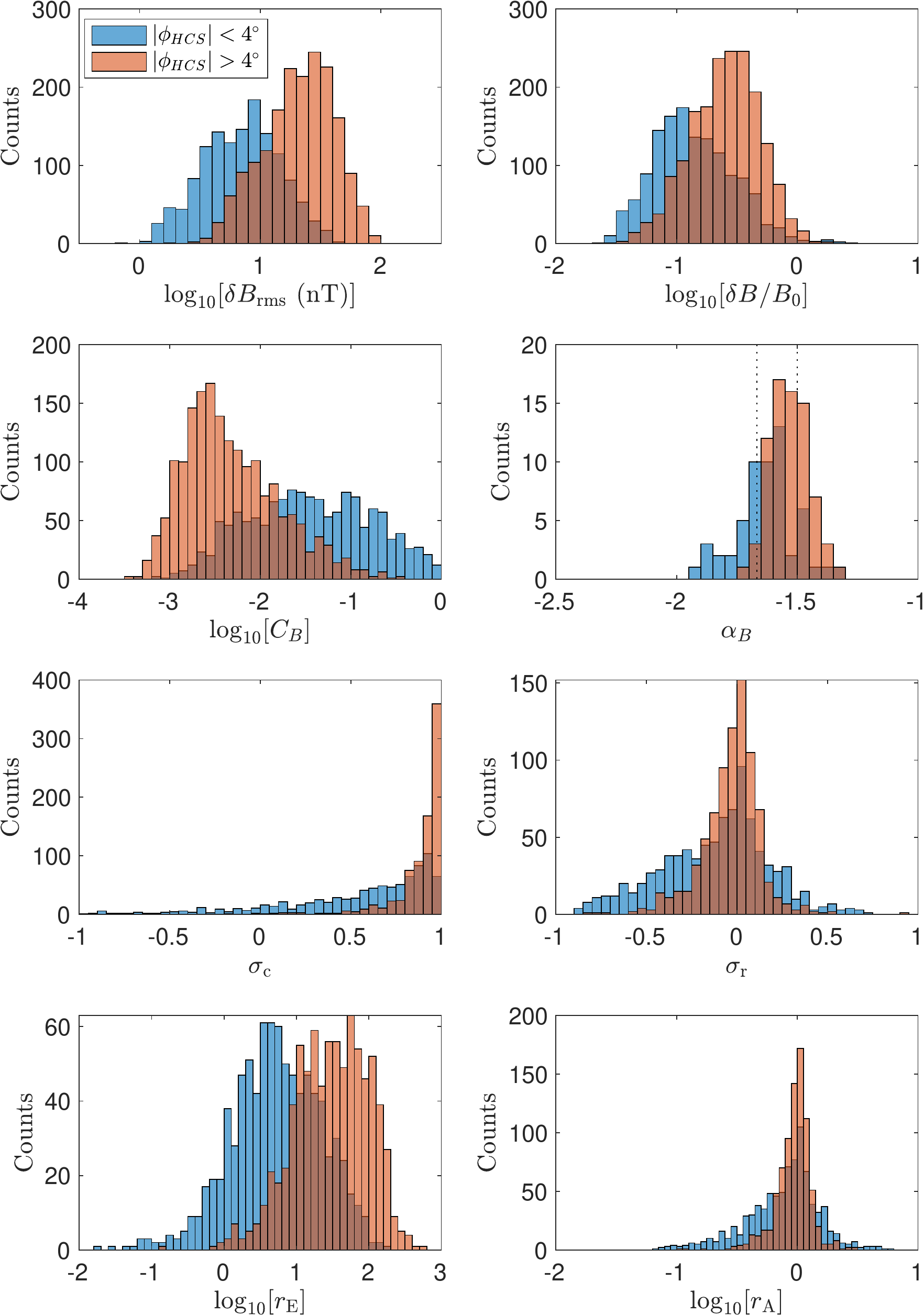}
\caption{Histograms of turbulence properties close to (blue) and far from (red) the HCS. A clear difference can be seen in all properties.}
\label{fig:histograms_hcs}
\end{figure}

A time series of these properties, along with the angular distance to the HCS, \phiHCS, is shown in Figure \ref{fig:turbulencetimeseries}. As expected, there is significant variability of all quantities, but there are also consistent trends over the encounter. The outbound portion appears to have lower fluctuations amplitudes, higher magnetic compressibility, a steeper spectral index, and be less dominated by pure outward Alfv\'enic fluctuations (\sigmac\ is closer to zero and \sigmar\ further from zero). This does not appear to be a consequence of radial distance (since the orbit is geometrically symmetric) or the angle of the magnetic field: the distribution of $\theta_\mathrm{BR}$ (reflected to lie in the range $0^\circ$ to $90^\circ$) is the same to within uncertainties, with a mean value of $\theta_\mathrm{BR}=26.8^\circ\pm0.4^\circ$ in both cases. One key difference, however, is the proximity to the HCS, the effect of which is explored in the rest of this Letter.

One important consideration is whether the Taylor hypothesis remains valid as \emph{PSP} gets closer to the Sun \citep{klein15a,bourouaine18,bourouaine19,bourouaine20a}. Figure \ref{fig:turbulencetimeseries} also contains the time series of the parameter $\epsilon=\delta v_\mathrm{rms}/\sqrt{2}v_\mathrm{sc}$ calculated from 1\,hour intervals, where $v_\mathrm{sc}$ is the magnitude of the solar wind velocity in the spacecraft frame. This is the same parameter as in the model of \citet{bourouaine19}, in which perpendicular sampling is assumed so $v_\mathrm{sc}\sim v_\mathrm{sc\perp}$, and in which the Taylor hypothesis is valid for $\epsilon\ll1$. The model also assumes Gaussian random sweeping and anisotropic turbulence $k_\perp\gg k_\|$, and is valid when $\tan(\theta_\mathrm{BV})\gtrsim\delta v_\mathrm{rms}/v_\mathrm{A}$, where $\theta_\mathrm{BV}$ is the angle between $\mathbf{B}_0$ and the mean solar wind velocity in the spacecraft frame. Points for which the model is valid are marked with blue dots and points for which it is not are marked as red crosses. \citet{bourouaine20a} determined that within this model, frequency broadening caused by the breakdown of the Taylor hypothesis does not significantly modify the spectrum as long as $\epsilon\lesssim0.5$, and as shown in Figure \ref{fig:turbulencetimeseries}, the data points satisfy this condition. Therefore, the differences in turbulence characteristics investigated in this Letter are likely not due to the differences in the validity of the Taylor hypothesis. A more detailed analysis of the Taylor hypothesis for these first \emph{PSP} orbits is given in \citet{perez20}.

\subsection{HCS Proximity Dependence}

\begin{figure}
\centering
\includegraphics[width=\columnwidth,trim=0 0 0 0,clip]{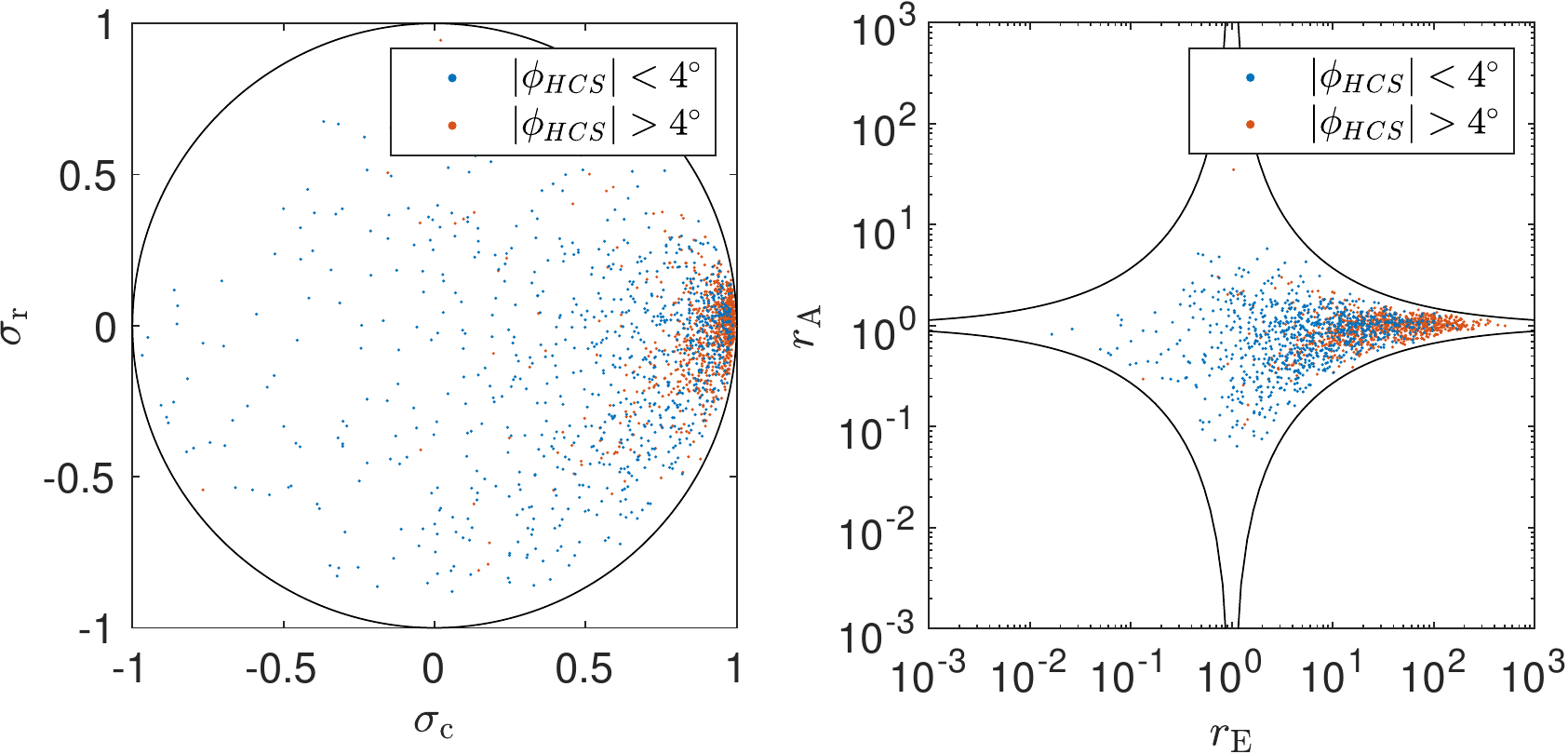}
\caption{Distributions of normalised cross helicity, \sigmac, normalised residual energy, \sigmar, Elsasser ratio, \rE, and Alfv\'en ratio, \rA, for times close to (blue) and far from (red) the HCS.}
\label{fig:circlestar}
\end{figure}

In the outbound portion of Encounter 4, \emph{PSP} spent significant time in the streamer belt wind near the HCS \citep{bale20}. The width of the streamer belt wind at these distances is not well known, so the dependence of the turbulence properties on the distance to the HCS was investigated. Figure \ref{fig:cutvalues} shows average values close to and far from the HCS, as a function of the cut value of the HCS angle used to define close and far, $\phi_\mathrm{cut}$. For example, the first panel shows $\left<\delta B_\mathrm{rms}\right>_{|\phiHCS|<\phi_\mathrm{cut}}$ as a function of $\phi_\mathrm{cut}$ in blue and $\left<\delta B_\mathrm{rms}\right>_{|\phiHCS|>\phi_\mathrm{cut}}$ as a function of $\phi_\mathrm{cut}$ in red. Because the imbalance is so high, plots for the Elsasser ratio,
\begin{equation}
\rE=\frac{1+\sigmac}{1-\sigmac},
\end{equation}
and Alfv\'en ratio,
\begin{equation}
\rA=\frac{1+\sigmar}{1-\sigmar},
\end{equation}
are also shown. All quantities show a difference at all cut angles, but the largest overall difference is between $\approx3^\circ$ and $\approx5^\circ$, so a value of $\phi_\mathrm{cut}=4^\circ$ was used to define the width of the region near the HCS in which the turbulence properties are different.

Figure \ref{fig:histograms_hcs} shows the distributions of the turbulence properties over the encounter both near to ($|\phiHCS|<4^\circ$) and far from ($|\phiHCS|>4^\circ$) the HCS. A clear difference can be seen in each property: near the HCS there are lower amplitudes, higher magnetic compressibility, a steeper spectrum, a lower level of imbalance and a broader distribution of residual energy. The joint distributions of \sigmac\ with \sigmar\ and \rE\ with \rA\ are shown in Figure \ref{fig:circlestar}. The data are constrained mathematically to lie within the regions 
\begin{equation}
\sigma_\mathrm{c}^2+\sigma_\mathrm{r}^2\leq1
\end{equation}
and
\begin{equation}
\left(\frac{\rE-1}{\rE+1}\right)^2+\left(\frac{\rA-1}{\rA+1}\right)^2\leq1,
\end{equation}
respectively, marked as solid black lines. In general, it can be seen that in both cases the fluctuations are highly Alfv\'enic ($\sigmac\approx1$, $\sigmar\approx0$, $\rE\gg1$, $\rA\approx1$), more so than in previous encounters \citep{chen20,mcmanus20,parashar20} but the near-HCS wind is less so than the wind far from the HCS.

\begin{table}
\caption[]{Solar wind and turbulence properties close to ($|\phiHCS|<4^\circ$) and far from ($|\phiHCS|>4^\circ$) the heliospheric current sheet (HCS). Quantities are arithmetic means, apart from those marked with $^*$, which are geometric means.}
\label{tab:properties}
$$
\begin{array}{ccc}
\hline
\noalign{\smallskip}
\mathrm{Property} & |\phiHCS|<4^\circ & |\phiHCS|>4^\circ \\
\noalign{\smallskip}
\hline
\noalign{\smallskip}
B\ \mathrm{(nT)} & 56 & 88  \\
n_\mathrm{e}\ \mathrm{(cm^{-3})} & 510 & 390  \\
v_\mathrm{i}\ \mathrm{(km\,s^{-1})} & 240 & 300  \\
T_\mathrm{i}\ \mathrm{(eV)} & 20 & 55  \\
4\pi\rho v_\mathrm{r}r^2\ (10^{-14}\,M_\odot\,\mathrm{yr}^{-1}) & 3.1 & 2.1  \\
2\pi\rho v_\mathrm{r}^3r^2\ \mathrm{(10^{19}\,W)} & 6.0 & 6.0  \\
\beta_\mathrm{i}^* & 1.2 & 0.68  \\
\noalign{\smallskip}
\hline
\noalign{\smallskip}
\delta B^*\ \mathrm{(nT)} & 7.0 & 21  \\
(\delta B/B_0)^* & 0.14 & 0.25  \\
C_B^* & 0.036 & 0.0048  \\
\alpha & -1.63 & -1.53  \\
\sigmac & 0.55 & 0.88  \\
\sigmar & -0.12 & -0.031  \\
\rE^* & 5.2 & 30  \\
\rA^* & 0.76 & 0.94  \\
f_\mathrm{b}\ \mathrm{(Hz)} & 3\times10^{-4} & 4\times10^{-3}  \\
\noalign{\smallskip}
\hline
\end{array}
$$
\end{table}

The averages for these different regions are given in Table \ref{tab:properties}. Aspects to note are that near the HCS the wind is denser, slower, cooler, with a higher plasma beta and mass flux, as expected for the streamer belt wind. The kinetic energy flux, however, is very similar, as has been seen previously across different wind types \citep{lechat12}. The difference in turbulence properties is consistent with the differences between ``fast'' and ``slow'' wind seen further from the Sun \citep[e.g.,][]{tu95,bruno13}. Notably, the magnetic field spectral index in the streamer belt wind is close to $-5/3$ whereas it is closer to $-3/2$ far from the HCS. This is consistent with observed dependencies of the magnetic spectrum on the degree of Alfv\'enicity at 1 au \citep{podesta10d,chen13b,bowen18} and in the previous \emph{PSP} orbits \citep{chen20}. A similar difference of spectral index and Alfv\'enicity was also seen when separating times of inverted and non-inverted magnetic field \citep{bourouaine20b}. The imbalance, however, is larger than typically seen at 1\,au, suggesting that the evolution towards a more balanced state seen in the open field wind \citep{chen20} also occurs in the streamer belt wind.

\subsection{Spectra}

\begin{figure}
\centering
\includegraphics[width=\columnwidth,trim=0 0 0 0,clip]{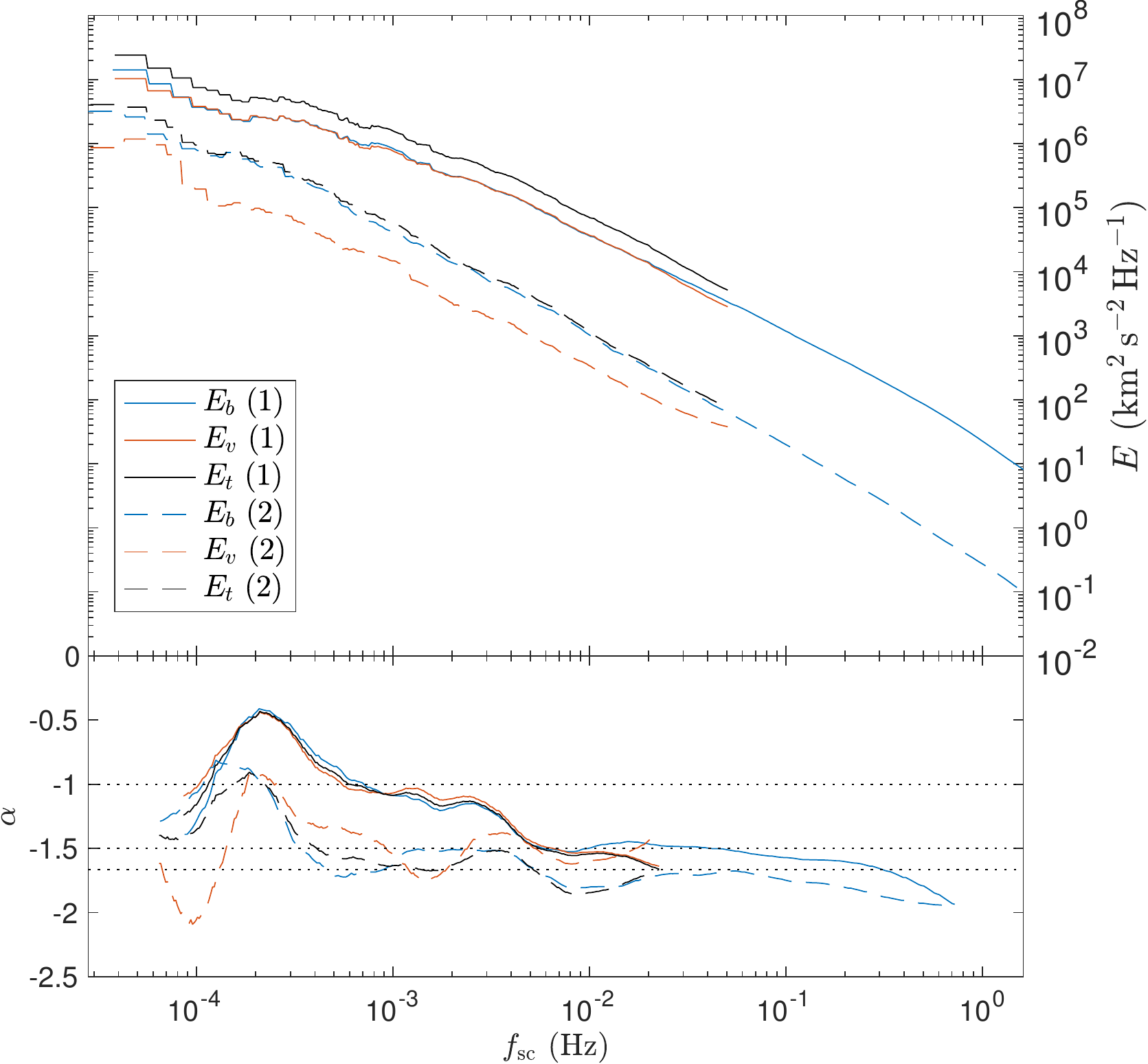}
\caption{Trace power spectra of magnetic, velocity, and total energy fluctuations ($E_b$, $E_v$, and $E_t$) for intervals 1 and 2 (indicated in Figures \ref{fig:timeseries} and \ref{fig:turbulencetimeseries}). The lower panel shows the local spectral index, $\alpha$, calculated over a sliding window of a factor of 5.}
\label{fig:spectra}
\end{figure}

Spectra of representative intervals (marked by the vertical dashed lines in Figures \ref{fig:timeseries} and \ref{fig:turbulencetimeseries}) are shown in Figure \ref{fig:spectra}. The lower panel shows the local spectral index, in which some important differences can be seen. Firstly, the steeper magnetic field spectrum in the near-HCS interval can be seen throughout the MHD inertial range. But more significantly, the break between the ``$1/f$'' range and the MHD inertial range is at very different frequencies: $f_\mathrm{b}\approx3\times10^{-4}$\,Hz in the near-HCS wind and $f_\mathrm{b}\approx4\times10^{-3}$\,Hz far from the HCS. The origin of the $1/f$ break is debated \citep{matthaeus86,velli89,verdini12,perez13,wicks13b,wicks13c,chandran18,matteini18}, but this difference is also seen between the ``fast'' and ``slow'' wind at 1 au \citep{bruno19}, where both breaks are about a decade lower in frequency, consistent with the radial evolution \citep{chen20}.

\subsection{Energy Flux}

\begin{figure}
\centering
\includegraphics[width=\columnwidth,trim=0 0 0 0,clip]{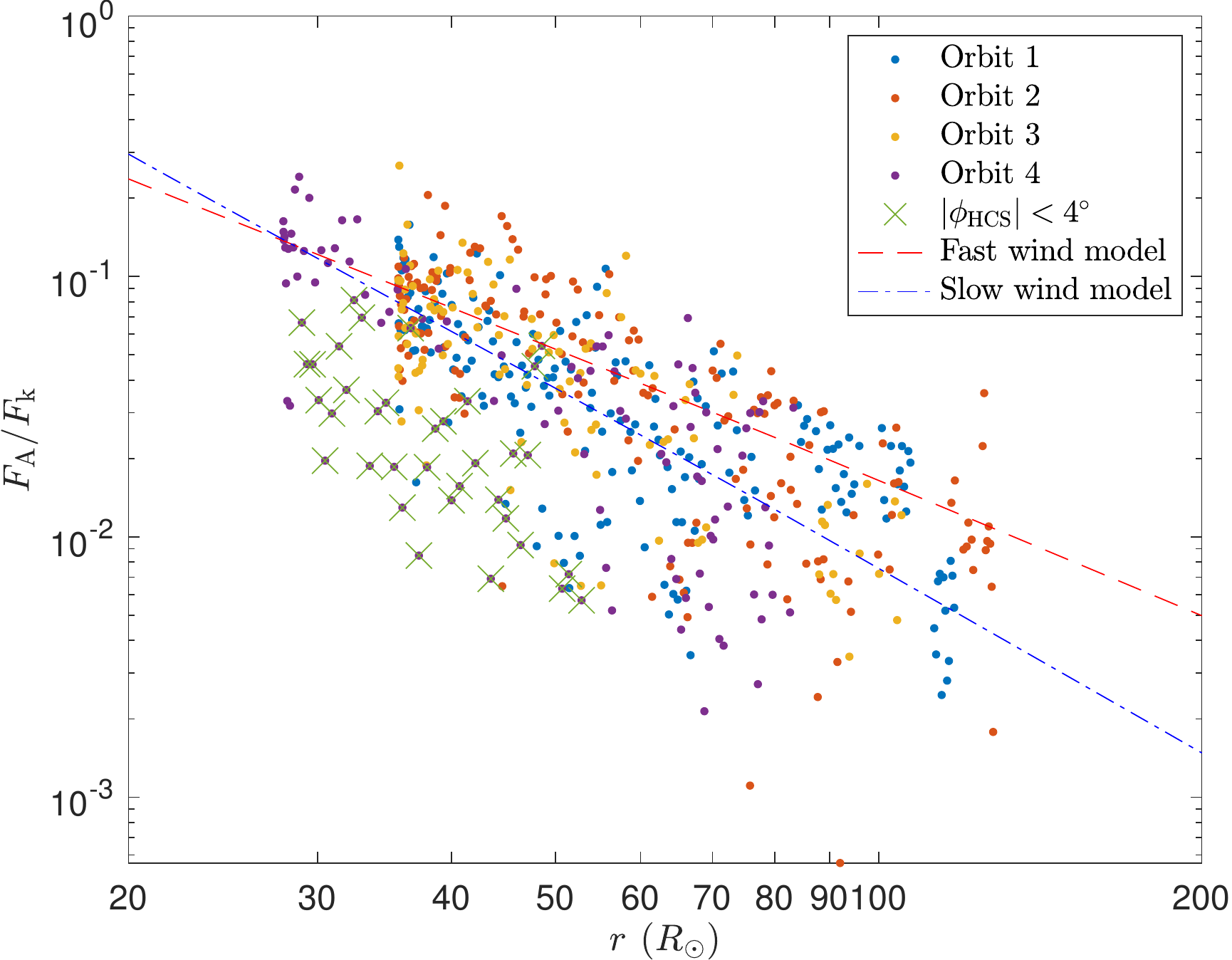}
\caption{Ratio of Alfv\'en wave energy flux, $F_\mathrm{A}$, to bulk solar wind kinetic energy flux, $F_\mathrm{k}$, for Orbits 1-4. Times during Orbit 4 which are close to the HCS ($|\phiHCS|<4^\circ$) are marked with green crosses. The fast and slow wind solutions to the \citet{chandran11} model are shown with the red dashed and blue dash-dotted lines.}
\label{fig:flux}
\end{figure}

Finally, the energy flux in the turbulent fluctuations was compared to the \citet{chandran11} reflection-driven turbulence solar wind model, following \citet{chen20}. The Alfv\'enic turbulence enthalpy flux was calculated as
\begin{equation}
F_\mathrm{A}=\rho|\delta\mathrm{b}|^2\left(\frac{3}{2}v_\mathrm{r}+v_\mathrm{A}\right),
\end{equation}
where $v_\mathrm{A}=B_0/\sqrt{\mu_0\rho_0}$ is the Alfv\'en speed, and the solar wind bulk kinetic energy flux as
\begin{equation}
F_\mathrm{k}=\frac{1}{2}\rho v_\mathrm{r}^3.
\end{equation}
These quantities are similar to those plotted in \citet{chandran11} and \citet{chen20} except $\delta\mathbf{b}$ is used instead of $\delta\mathbf{z}^+$, which allows greater data coverage and is valid due to the high degree of Alfv\'enicity (Figure \ref{fig:circlestar}). Figure \ref{fig:flux} shows the ratio $F_\mathrm{A}/F_\mathrm{k}$, calculated over 3 hour intervals, as a function of solar distance over the first four orbits of \emph{PSP}. The SPAN-I data were used for $v_\mathrm{r}$ for times when reliable fits could be made, and SPC data used otherwise. Also shown are two solutions to the \citet{chandran11} model as described in \citet{chen20}.

It can be seen that for Orbits 1-3, the data, on average, follow the model solutions (although with some degree of spread), consistent with the findings of \citet{chen20}. For Orbit 4, however, there are a large number of points that fall well below the model, originating from times during the outbound portion of the encounter near the HCS when the turbulent amplitudes were lower. Times when the spacecraft was close to the HCS during Encounter 4 ($r<54\,R_\odot$) are marked with green crosses. These values are on average 3.0 times smaller than the slow wind model (whereas the average ratio to the model for the other data points over this distance range is 1.0). This suggests that the lower amplitude turbulence near the HCS does not contain sufficient energy for reflection-driven Alfv\'enic turbulence models to provide an explanation for the acceleration of the streamer belt wind.

\section{Discussion}

In this Letter, it is shown that the turbulence properties in the near-Sun streamer belt wind, from $\approx28\,\RS$ to $\approx54\,\RS$, are significantly different to in the open field wind that has been measured for most of the previous \emph{PSP} orbits. These differences include lower amplitudes, higher magnetic compressibility, a steeper magnetic spectrum (--5/3 rather than --3/2), a lower degree of Alfv\'enicity, and a larger scale $1/f$ break. The differences are similar to the traditional fast/slow wind differences reported at 1 au, suggesting the near-Sun streamer belt wind as the prototypical slow solar wind.

The differences in turbulence properties occur at an angle to the HCS of $|\phiHCS|\approx4^\circ$, suggesting the total width of the streamer belt wind at these solar distances to be $\approx8^\circ$. This interpretation is consistent with other studies of the streamer belt wind during PSP Encounter 4, e.g., \citet{badman20b} found a reduced magnetic flux during this period that did not fit the general radial scalings seen in the rest of the PSP data. The inferred streamer belt wind width is also consistent with coronal images that show the streamer rays to have a width of around $10^\circ$ to $20^\circ$, which is likely a slight over-estimate due to the line-of-sight integration \citep{rouillard20,poirier20}. 

The Alfv\'enic turbulence energy flux measured on \emph{PSP}'s first three orbits is generally in line with that required for the reflection-driven solar wind model of \citet{chandran11}, however, in the Encounter 4 streamer belt wind it is several times lower than the model predictions. There are also occasional periods in the previous orbits where this is the case, identified by \citet{chen20} as the periods of quiet radial-field wind seen by \citet{bale19}. This raises the possibility that these may be small patches of streamer belt wind; investigating this possibility would be an interesting topic for further study. The difference to the model predictions implies that purely reflection-driven Alfv\'enic turbulence solar wind models \citep[e.g.,][]{cranmer07,chandran11} may not be able to account for the acceleration of the streamer belt wind, and additional processes are taking place. One possible such processes is additional turbulence driving by velocity shears \citep{usmanov18,ruffolo20}. One thing to note about this possibility, however, is that it would be expected to produce both inward and outward Alfv\'en waves, resulting in $\sigmac\approx0$ and $\rE\approx1$ if the shears dominate the energy input. In the streamer belt wind measured here, $\sigmac=0.55$ and $\rE=5.2$ on average, meaning that even though the imbalance is less than in the open field wind, there is still 5 times more energy in the outward waves compared to the inward ones, so such processes may play a role but cannot be dominating the energy input overall. Another likely contribution to the generation of the streamer belt wind is reconnection in the near-Sun HCS \citep{lavraud20} triggered by a tearing mode \citep{reville20b}. The chain of processes involved in the acceleration of the streamer belt wind, however, remains an open question. Future orbits of \emph{PSP} closer to the Sun will allow more to be learnt about the nature of both plasma turbulence near the Sun and solar wind acceleration.

\begin{acknowledgements}
CHKC was supported by STFC Ernest Rutherford Fellowship ST/N003748/2 and STFC Consolidated Grant ST/T00018X/1. LDW was supported by STFC Consolidated Grant ST/S000364/1.  JCP was partially supported by NASA grants NNX16AH92G, 80NSSC19K0275 and NSF grant AGS-1752827. SB was supported by NASA grants NNX16AH92G, 80NSSC19K0275 and 80NSSC19K1390. KGK was supported by NASA ECIP grant 80NSSC19K0912. SDB was supported in part by the Leverhulme Trust Visiting Professorship programme. We thank the members of the FIELDS/SWEAP teams and \emph{PSP} community for helpful discussions. \emph{PSP} data is available at the SPDF ({https://spdf.gsfc.nasa.gov}).
\end{acknowledgements}

\bibliographystyle{aa}
\bibliography{bibliography} 

\end{document}